\documentclass[a4,aps,prb,twocolumn,superscriptaddress,nofootinbib]{revtex4}
\usepackage{amsmath}
\DeclareMathOperator{\Tr}{Tr}
\usepackage{graphicx}
\usepackage{latexsym}
\usepackage{array}
\usepackage{color}
\newcolumntype{P}[1]{>{\centering\arraybackslash}p{#1}}
\newcolumntype{M}[1]{>{\centering\arraybackslash}m{#1}}
\input{epsf}
\begin{document}
\title{Elimination of thermal bistability in superconducting weak links by an inductive shunt}
\author{Sourav Biswas}
\affiliation{Department of Physics, Indian Institute of Technology Kanpur, Kanpur 208016, India}
\author{Clemens B. Winkelmann}
\affiliation{\mbox{Univ.} Grenoble Alpes, CNRS, Grenoble INP, Institut N\'eel, Grenoble, France}
\author{Herv\'e Courtois}
\affiliation{\mbox{Univ.} Grenoble Alpes, CNRS, Grenoble INP, Institut N\'eel, Grenoble, France}
\author{Thierry Dauxois}
\affiliation{\mbox{Univ.} Lyon, ENS de Lyon, \mbox{Univ.} Claude Bernard, CNRS, Laboratoire de Physique, Lyon, France}
\author{Hillol Biswas}
\affiliation{Department of Physics, Indian Institute of Technology Kanpur, Kanpur 208016, India}
\author{Anjan K. Gupta}
\email[]{anjankg@iitk.ac.in}
\affiliation{Department of Physics, Indian Institute of Technology Kanpur, Kanpur 208016, India}
\date{\today}

\begin{abstract}
The quantum phase-coherent behavior of superconducting weak links (WL) is often quenched in the finite voltage state, due to the heat dissipation and related thermal hysteresis. The latter can be reduced by improving heat evacuation and/or by lowering the critical current, so that a phase-dynamic regime is obtained, albeit over a narrow bias-current and temperature range. Here we demonstrate that an inductive shunt with well-chosen parameters introduces unexpected nonlinear dynamics that destabilize an otherwise stable fixed point in the dissipative branch. This leads to a nonhysteretic behavior with large voltage oscillations in intrinsically hysteretic WL-based micron-size superconducting quantum interference devices. A dynamic thermal model quantitatively describes our observations and further allows us to elaborate on the optimal shunting conditions.
\end{abstract}
\maketitle

\section{Introduction}
Superconducting weak links (WL) \cite{likharev} acting as Josephson junctions are of great interest for a range of quantum applications. A WL can be probed with a \mbox{d.c.} current bias in the phase dynamic state \cite{tinkhambook} so that a \mbox{d.c.} voltage is measured. In particular, a WL-based micron-size superconducting quantum interference device ($\mu$-SQUID) features then a flux-sensitive voltage.\cite{sourav} A magnetic moment resolution better than 1 $\mu_{\rm B}$ can be reached,\cite{zeldov} which makes it an ultimate probe for quantum nanomagnetism.\cite{recentsquidreview,granatasquid,mag2-werns,highqualitysquid,moler} The main limitation to $\mu$-SQUIDs operation resides in the (thermal) hysteresis of their current-voltage characteristics (IVCs) at low temperatures, due to large critical current and poor heat evacuation from the WL to the bath.\cite{herveprl,scocpol,tinkhamnanowire,nikhilprl} A time-dependent Ginzburg-Landau approach capturing non-equilibrium effects on the order-parameter relaxation can model the hysteresis and the phase-dynamic regime in WLs.\cite{tdgl,peeters,peeters2} A simpler dynamic thermal model (DTM) successfully describes the same behavior by considering both the phase dynamics and the Joule heat evacuation.\cite{anjanjap,sourav} This gives scope for further optimization so as to obtain a phase-dynamic regime as a mono-stable state over a wide bias current and temperature range. In particular, a resistive shunt \cite{baranovprb,nikhilsust,bezryadin} placed close to a WL can remove thermal hysteresis down to a certain temperature. However, for very low temperatures the required small resistor makes the $\mu$-SQUID voltage modulation minuscule. Shunts with a larger inductance lead to relaxation oscillations \cite{relaxation} due to the induced delay in current switching. Shunts with intermediate inductance provide a wider parameter space, \cite{bellsustview} which is not yet investigated, for optimizing WL based devices.

In this paper, we uncover the striking effect of fine-tuned inductive shunting on the behavior of a WL-based $\mu$-SQUID. When we use a shunt made of a resistor with an adequate inductor in series, we observe reversible IVCs with a large voltage modulation by the magnetic flux over an increased temperature and bias current range. The dynamic retrapping current increases with the inductance up to a limiting value above which relaxation oscillations appear. The dynamic thermal model incorporating the nonlinear dynamics of temperature and current quantitatively explains the observations.

\section{Effect of Inductive Shunt on Phase Dynamics: A Model}
Across a WL in a non-zero voltage state, superconducting phase-correlations can remain, so that the bias current through it is dynamically shared between a normal current and a superconducting component. Due to dissipation from periodic phase slips, the WL heats up above the bath temperature $T_{\rm b}$. Still it can remain at a temperature $T_{\rm WL}$ below the WL critical temperature $T_{\rm c}$,\cite{anjanjap} provided the heat conduction to the bath is efficient enough. In this case, the Josephson coupling persists. The occurrence of this dissipative regime, called dynamic regime, defines a current range between the {\it dynamic} retrapping current $I_{\rm r}^{\rm dyn}$ and the {\it static} retrapping current $I_{\rm h}$. For $I<I_{\rm r}^{\rm dyn}$, the dynamic state is not  possible, leading the WL to the superconducting, zero-voltage state. For $I >I_{\rm h}$, the WL temperature $T_{\rm WL}$ exceeds $T_{\rm c}$, leading to the loss of any phase correlation across the WL.

In the DTM, the thermal heat loss from the WL to the substrate is described by $k(T_{\rm WL}-T_{\rm b})$. A dimensionless parameter\cite{anjanjap}
\begin{align}
\beta =\frac{{I_{\rm c}^{\rm 0}}^2(T_{\rm b})R_{\rm N}}{k(T_{\rm c}-T_{\rm b})}
\end{align}
then determines the accessibility of the dynamic regime at a given bath temperature $T_{\rm b}$. Here $I_{\rm c}^{\rm 0}$, $R_{\rm N}$ and $k$ are the zero-field critical current, normal resistance and heat loss coefficient of the WL, respectively. The two extreme values $\beta \to 0$ and $\beta \gg 1$ lead the DTM to the isothermal RSJ \cite{squidbook,tinkhambook} and static thermal models,\cite{scocpol} respectively.
	
\begin{figure}[t!]
\includegraphics[width=0.8\columnwidth]{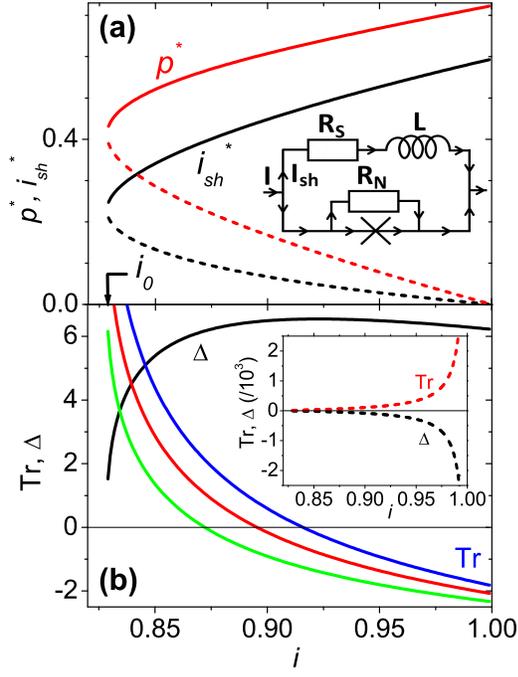}
\caption{Calculations for the case of $\beta$ = 6 and $r$ = 2. (a) The first (dashed lines, unstable) and the second (continuous lines) fixed point coordinates $p^*$, $i_{\rm sh}^*$ as a function of the bias current $i$. Inset: equivalent circuit diagram of a WL shunted by a resistor $R_{\rm S}$ and inductor $L$. (b) Variation of the determinant $\Delta$ (black) and Trace Tr at $\gamma/\alpha=$ 1.1 (green), 1.3 (red) and 1.5 (blue), with the bias current $i$ for the second fixed point. The inset shows the $\Delta$ and Tr variation for the first fixed point.}
\label{fig:model1}
\end{figure}

We consider a WL, with normal resistance $R_{\rm N}$, that is resistively and inductively shunted with a resistance $R_{\rm S}$ and an inductance $L$ in series, as shown in \mbox{Fig. \ref{fig:model1}(a)} inset. The time evolution of the shunt current $I_{\rm sh}$ is described by the equation:
\begin{align}
L\frac{dI_{\rm sh}}{dt}+I_{\rm sh}R_{\rm S}=V=\frac{\Phi_{\rm 0}}{2\pi}\frac{d\varphi}{dt},
\end{align}
with V and $\varphi$ as the voltage and the phase difference across the WL, respectively. Writing, in addition, an RSJ-type equation and the heat balance in the WL, one obtains the full set of dimensionless equations determining the dynamics of phase, temperature and shunt current:\cite{anjanjap,sourav}
\begin{align}
\dot{\phi} &= i-(1-p)\sin(2\pi\gamma\phi)-i_{\rm sh} \label{eq:eq1} \\
\dot{p} &=-\frac{\gamma}{\alpha}p+\beta\frac{\gamma}{\alpha}\dot{\phi}^{\rm 2} \label{eq:eq2} \\
\dot{i_{\rm sh}} &=-i_{\rm sh} + r\dot{\phi}.
\label{eq:eq3}
\end{align}
The relevant time scales are the thermal time $\tau_{\rm th} = C_{\rm WL}/k$, the Josephson time $\tau_{\rm J}=\Phi_{\rm 0}/I_{\rm c}^{\rm 0}(T_{\rm b})R_{\rm N}$, and the inductive time $\tau_{\rm L}=L/R_{\rm S}$. Here, $C_{\rm WL}$ is the WL heat capacity and $I_{\rm c}^{\rm 0}(T_{\rm b})$ is assumed to vary linearly with $T_{\rm b}$. We introduce the parameters $r=R_{\rm N}/R_{\rm S}$, $\gamma=\tau_{\rm L}/\tau_{\rm J}$ and $\alpha = \tau_{\rm th}/\tau_{\rm J}$. The time unit is $\tau=t/(\gamma\tau_{\rm J})$ and we use the reduced phase $\phi=\varphi/(2\pi\gamma)$ and the reduced temperature  $p = (T_{\rm WL}- T_{\rm b})/(T_{\rm c}-T_{\rm b})$. Currents denoted by $i$ with relevant sub/super-script represent the same in units of $I_{\rm c}^{\rm 0}(T_{\rm b})$.

We focus our analysis on the large $\alpha, \gamma$ limit that is relevant in most practical cases. In this limit, the time evolution of the phase $\phi$ is much faster than that of the temperature $p$ and the shunt current $i_{\rm sh}$. Therefore, the deviation in $p$ and $i_{\rm sh}$, over the phase-slip time $\tau_{\rm ps}$, from their time averages $\overline{p}$ and $\overline{i}_{\rm sh}$ can be neglected. Here, $\tau_{\rm ps}$ is the time over which the phase $\varphi$ changes by $2\pi$. By integrating \mbox{Eq. (\ref{eq:eq1})} over this $\tau_{\rm ps}$, one gets:
\begin{align}
\tau_{\rm ps}=2\pi/\sqrt{(i-\overline{i}_{\rm sh})^2-(1-\overline{p})^2}.
\label{eq:time}
\end{align}
The averages $\overline{\dot{\phi}^2}$ and $\overline{\dot{\phi}}$ are obtained as $\overline{\dot{\phi}^2}=2\pi(i-\overline{i}_{\rm sh})/\tau_{\rm ps}$ and $\overline{\dot{\phi}}=2\pi/\tau_{\rm ps}$. Taking the time-averages of \mbox {Eqs. (\ref{eq:eq2}) and (\ref{eq:eq3})} over $\tau_{\rm ps}$, we thus obtain a two-dimensional dynamical system:
\begin{align}
\frac{\alpha}{\gamma}\,\overline{\dot{p}} &= -\overline{p}+\beta(i-\overline{i}_{\rm sh})\sqrt{(i-\overline{i}_{\rm sh})^2-(1-\overline{p})^2} \label{eq:eq4}\\
\overline{\dot{i_{\rm sh}}} &= -\overline{i}_{\rm sh}+r\sqrt{(i-\overline{i}_{\rm sh})^2-(1-\overline{p})^2}. \label{eq:eq5}
\end{align}
In this set of two non-linear equations, the ratio $\gamma/\alpha$ determines the dynamics of the temperature $p$, and hence of the WL critical current $i_{\rm c}(p)=1-p$. In contrast, the dynamics of the shunt current $i_{\rm sh}$ is not (directly) dependent on $\gamma/\alpha$. Moreover, Eqs. (\ref{eq:eq4}) and (\ref{eq:eq5}) do not correspond to any well-defined (conservative) potential like the tilted washboard potential in an isothermal RCSJ model\cite{squidbook} or the fictitious potential in the DTM.\cite{anjanjap} We are thus led to pursue an analysis involving the fixed points of the system and their stability.

Removing the average symbol for the sake of simplicity, one can write two relations between the fixed points' coordinates $p^*$ and $i_{\rm sh}^*$ with the bias current $i$ as a parameter:
\begin{align}
p^*&=\beta(i-i_{\rm sh}^*)i_{\rm sh}^*/r \label{eq:eq6}  \\
p^*&=1-\sqrt{(i-i_{\rm sh}^*)^2-(i_{\rm sh}^*)^2/r^2}. \label{eq:eq7}
\end{align}
As elaborated in Appendix A, Equations (\ref{eq:eq6},\ref{eq:eq7}) feature two real and non-zero solutions ($p^*$,$i_{\rm sh}^*$) only when the bias current $i$ is above a threshold $i_0$, where a saddle-node bifurcation occurs.\cite{nonlinear-book} This threshold, as well as the fixed points' coordinates, depend on $\beta$ and $r$ but not on $\alpha$ or $\gamma$. \mbox{Figure \ref{fig:model1}(a)} shows the variation of the coordinates ($p^*$,$i_{\rm sh}^*$) with the bias current $i$ for $\beta=6$ and $r=2$. The considered $\beta$ value is large as our main interest is in intrinsically deeply hysteretic $\mu$-SQUIDs.\cite{sourav} As elaborated further in the following, the first fixed point with the lower (but non-zero) values of $p^*$ and $i_{\rm sh}^*$ is found to be always unstable, while the second one can be stable or unstable.

Our focus here is on finding the values of the bias current $i$ for which a stable, non-zero fixed point exists. Here, these values cover a continuous bias current range. We identify the lower limit of this range as the dynamic retrapping current $i_{\rm r}^{\rm dyn}$. In the absence of fluctuations, a WL in the finite voltage state and with a decreasing bias current would retrap to the superconducting state precisely at $i=i_{\rm r}^{\rm dyn}$.

The stability of the fixed point is dictated by the trace $\Tr$ and determinant $\Delta$ of the Jacobian matrix associated with the dynamical system: a stable fixed point requires $\Delta>0$ and $\Tr<0$.\cite{nonlinear-book} Using \mbox{Eqs.} (\ref{eq:eq4}) and (\ref{eq:eq5}), one obtains:
\begin{align}
\Tr & =\frac{r^2p^*}{(i_{\rm sh}^*)^2}\left[\frac{\gamma}{\alpha}(1-p^*)-\frac{r}{\beta}\right]-1-\frac{\gamma}{\alpha} \label{eq:eq8} \\
\Delta & =\frac{\gamma}{\alpha}\left[1+\frac{r^3p^*}{\beta(i_{\rm sh}^*)^2}+r(1-p^*)\left\{\beta-\frac{rp^*}{(i_{\rm sh}^*)^2}\right\}\right]. \label{eq:eq9}
\end{align}
Whereas the fixed point coordinates as well as the sign of $\Delta$ are independent of $\gamma/\alpha$, the sign of $\Tr$, and hence the stability of the fixed points, depends on $\gamma/\alpha$.

\begin{figure}[t!]
\includegraphics[width=0.8\columnwidth]{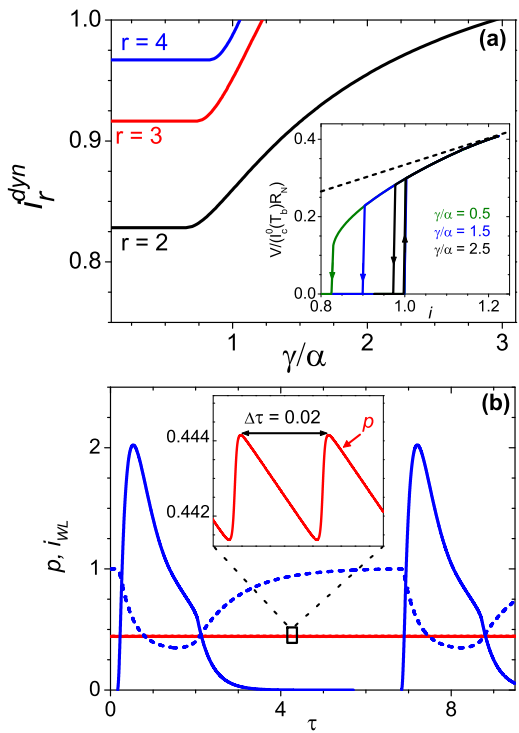}
\caption{(a) Reduced dynamic retrapping current $i_{\rm r}^{\rm dyn}$ variation with $\gamma/\alpha$ for fixed $\beta=6$ and  $r$ = 2, 3 and 4. Inset: IVCs for different $\gamma/\alpha$ values in the dynamic regime without relaxation oscillations for $r=2$. The dotted line represents the Ohmic branch. (b) Time trace of the reduced temperature $p$ (continuous lines) and WL current $i_{\rm WL}$ (dashed line) for $\beta$= 6 and $r$= 2. The inset shows small oscillations at $\gamma/\alpha=0.5$ (red lines, only p is shown). For $\gamma/\alpha=3$ (blue lines), $p$ and $i_{\rm WL}$ exhibit relaxation oscillations.}
\label{fig:model2}
\end{figure}
	
\mbox{Figure \ref{fig:model1}(b)} shows the trace and the determinant for the first fixed point (inset), and for the second one (main panel), again for $\beta=6$ and $r=2$. Owing to a negative determinant $\Delta$, the first fixed point is always unstable, irrespective of $i$ and $\gamma/\alpha$ values. Regarding the second fixed point, the determinant $\Delta$ is shown as a function of the bias current for $\gamma/\alpha=1.3$ while the trace $\Tr$ is shown for $\gamma/\alpha=1.1, 1.3$ and 1.5. The trace becomes negative, \mbox{i.e.} the second fixed point becomes stable, with increasing bias current. This crossover defines the dynamic retrapping current $i_{\rm r}^{\rm dyn}$, which is found to be larger than $i_0$. Note that this crossover is actually a Hopf bifurcation.\cite{nonlinear-book} The dynamic retrapping current $i_{\rm r}^{\rm dyn}$ is found to be very close to $i_0$ for $\gamma/\alpha$ values much less than one. With $\gamma/\alpha$ increasing above one, it rises monotonically. This behavior can be seen in \mbox{Fig. \ref{fig:model2}(a)} where $i_{\rm r}^{\rm dyn}$ rises from a plateau at $i_0$.  The inset shows the simulated IVCs obtained by taking a time-average of the phase derivative $d\phi/d\tau$. In agreement with the above discussion, one observes a reduction in the current range of the bistable regime when $\gamma/\alpha$ is increased from 0.5 to 2.5.

Therefore, when the inductive and thermal times (or $\gamma$ and $\alpha$) are close to each other, a large portion of the dissipative, phase-dynamic branch becomes unstable. A physical understanding for this could be as follows. Near the stable fixed point (with non-zero voltage), the WL temperature-dynamics gets perturbed periodically by the inductive current-dynamics having a similar time scale. The matching of these time scales, resembling a resonance-like condition, leads the system far enough from its stable fixed point to eventually end up with the other fixed point, \mbox{i.e.} the zero-voltage superconducting state. On the other hand, when $\gamma/\alpha$ is well below 1, the inductance has little influence on the shunt current dynamics (as it is much faster than the temperature dynamics) and consequently little contribution to the overall behavior. Thus the dynamic retrapping current remains close to the prediction of the zero-inductance DTM.

For the other limit of large $\gamma/\alpha$, the current switching from the WL to the shunt is slowed down by the large inductance, leading to a sharp rise in WL temperature above $T_{\rm c}$ when the bias current exceeds the WL critical current. Once enough current is diverted away from the WL over a time determined by inductance, the WL starts cooling and at some time its temperature goes below $T_{\rm c}$. This leads to a larger current through the WL and cooling accelerates as a part of the current flows as supercurrent. This trend gets interrupted when the WL current exceeds again the critical current, leading to repetition of the same cycle.\cite{relaxation} The WL current $i_{\rm WL}$($=i-i_{\rm sh}$) and the WL temperature $p$ then exhibit relaxation oscillations with a dramatic time-dependence, as shown in \mbox{Fig. \ref{fig:model2}(b)}.

To summarize this theoretical description, an inductive shunt brings in a new dynamical variable, \mbox{i.e.} the shunt current. The single parameter $\gamma/\alpha$ can destabilize the otherwise stable fixed point of the 2D nonlinear dynamical system constituted by the dissipative WL coupled to the thermal bath. Thus an appropriate inductive shunt can enhance the reversibility of a WL-based $\mu$-SQUID and enable large voltage modulations in the SQUID response.

\section{Experimental Details}

\begin{figure}[t!]
	\includegraphics[width=0.5\columnwidth]{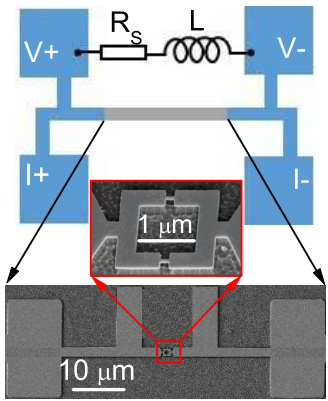}
	\caption{Schematic of a $\mu$-SQUID shunted by a resistor $R_{\rm S}$ and an inductor $L$. The zoomed-in portion shows large and small scale SEM images of the device.}
\label{fig:sem-schem}
\end{figure}
We fabricated $\mu$-SQUIDs on a silicon substrate using lift-off of an Al mask and Nb etch following a recipe discussed elsewhere.\cite{sourav} The length and width of the $\mu$-SQUID WLs are 160 nm and 40 nm respectively. Four probe transport measurements were performed using the setup as in Ref. \onlinecite{sourav} down to 1.3 K. The onset of superconductivity is seen at 8.6 K.

For a shunt resistance, a Nichrome wire was connected in parallel to the device's voltage leads and at a distance from the $\mu$-SQUID of about 1 cm. An estimate of the shunt loop inductance $L$ gives a few nH, already much larger than the total (geometric and kinetic) $\mu$-SQUID inductance $L_{\rm SQ}$, which is of pH order.\cite{nikhilsust} The screening parameter\cite{squidbook} $\beta_{\rm L}=L_{\rm SQ}I_{\rm c}^0/\Phi_{\rm 0}$ for our devices is less than 0.1. We used two different shunt resistance values $R_{\rm S1}$ = 4 $\Omega$ and $R_{\rm S2}$ = 2 $\Omega$. In that case, the inductive time $\tau_{\rm L}$ is of the order of few ns, the Josephson time $\tau_J$ is about 100 ps while $\tau_{\rm th}$ is of the order of a $\mu$s (see below) so that $\gamma/\alpha \simeq 10^{-3}$. In that regime, the inductance has little effect on the dynamic behavior.

Inductive shunts in the $\mu$H range, resulting in $\tau_{\rm L}$ of order $\mu s$ and $\gamma/\alpha \simeq 1$, were realized by a superconducting wire coil. The related magnetic flux coupled to the SQUID loop is estimated to be negligible compared to $\Phi_{\rm 0}$. The schematic of a $\mu$-SQUID, shunted by a resistor $R_{\rm S}$ and an inductor $L$ is shown in Fig. \ref{fig:sem-schem} along with a large scale SEM image and the SQUID loop.

\begin{figure}[t!]
\includegraphics[width=\columnwidth]{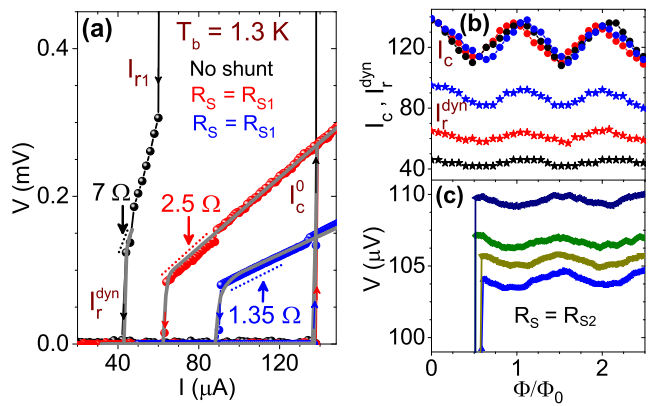}
\caption{Experimental results at 1.3 K. (a) Hysteretic IVCs at zero magnetic flux for three shunt cases. Solid gray lines are fits to the DTM (without inductance). (b) Modulations in $I_{\rm c}$ and $I_{\rm r}^{\rm dyn}$ with the flux $\Phi$. Color codes are same as panel (a). (c) $V(\Phi)$ modulation for the shunt $R_{\rm S2}$, displayed over the range $I = 108-112$ $\mu$A.}
\label{fig:ivc}
\end{figure}

In the following, we discuss results from a single device but with different shunting conditions. Similar results from another device can be found elsewhere.\cite{sourav-thesis} Figure \mbox{\ref{fig:ivc}(a)} shows the IVCs at 1.3 K and zero external magnetic flux. With no shunt, a strong hysteresis is seen with a critical current $I_{\rm c}^{\rm 0}\approx 137$ $\mu$A and a retrapping current $I_{\rm r}^{\rm dyn}\approx 42$ $\mu$A. A thermal instability\cite{nikhilprl} in the SQUID leads occurs above $I_{\rm r1}\approx$ 60 $\mu$A. The differential resistance dV/dI above $I_{\rm r}^{\rm dyn}$ is found to be 7 $\Omega$. The low dV/dI compared to the WL normal-state resistance and the modulation of $I_{\rm r}^{\rm dyn}$ with the magnetic flux $\Phi$ in \mbox{Fig. \ref{fig:ivc}(b)} confirm that the supercurrent is not completely destroyed in the dissipative state just above $I_{\rm r}^{\rm dyn}$.

With a resistive shunt, the dynamic retrapping current $I_{\rm r}^{\rm dyn}$ increases significantly while the critical current $I_{\rm c}^{\rm}$ remains the same. Still, the increased value of $I_{\rm r}^{\rm dyn}$ remains well below the $I_{\rm c}^{0}$ value, even with the lowest shunt resistor $R_{\rm S2}$. In the zero-inductance limit of relevance here, $\beta$ is the single parameter to describe the characteristics. Using the DTM \cite{sourav} and as elaborated in Appendix B, we obtain $\beta$(unshunted) = 9.3, $\beta(R_{\rm S1})$ = 4.2 and $\beta(R_{\rm S2})$ = 1.8 at 1.3 K. A practical SQUID operation in the dynamic regime, defined as $\beta<2$,\cite{sourav} is thus obtained over a wider temperature range if the device is resistively shunted. The shunted samples IVCs' slopes at large current lead to estimates of the shunt resistance of about 3.85 $\Omega$ and 1.67 $\Omega$, close to the measured values at room temperature.

The $V(\Phi)$ oscillations are observed down to 2.2 K for the unshunted device, and till 1.8 and 1.3 K (the lowest temperature investigated here) for the $R_{\rm S1}$ and $R_{\rm S2}$ shunted devices respectively, see \mbox{Fig. \ref{fig:ivc}(b)}. Nevertheless, resistive shunting neither improves modulation amplitude nor sensitivity in a significant way. The best flux noise density $\sqrt{S_{\Phi}}$ = 30 $\mu \Phi_{\rm 0}/\sqrt{\rm Hz}$ with the $R_{\rm S2}$ shunt is obtained at 2.2 K when the IVCs are nonhysteretic. Further detailed results on pure resistive shunting are presented in Appendix B.

\section{Effect of inductive shunt on $\mu$-SQUID}
\begin{figure}[t!]
	\includegraphics[width=\columnwidth]{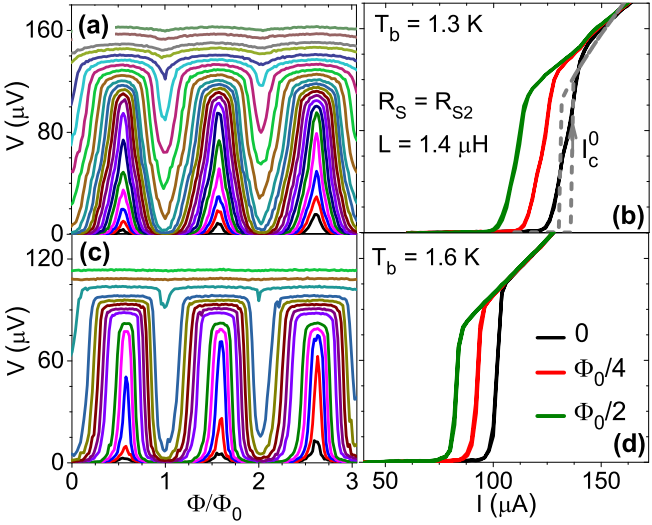}
	\caption{(a,c) $V(\Phi)$ modulations with an inductive shunt made of $R_{\rm S2}$ and $L$ = 1.4 $\mu$H at 1.3 and 1.6 K, respectively. The bias current ranges are respectively 110-150 $\mu$A and 80-112 $\mu$A. (b,d) IVCs at three different flux values at the respective temperatures. The gray dashed line in (b) represents the zero-field IVC calculated using the model with $\gamma/\alpha$ = 1.01, $\beta$= 9.3 and $r$= 4.2.}
	\label{fig:bvosc1}
\end{figure}

We now discuss experiments on the same device but with an inductive shunt, which is the main focus of this work. For the results discussed here, the shunt is made of the resistance $R_{\rm S2}$ and an inductance $L$ whose value was varied by changing the winding. For $L$ below 1 $\mu$H, no change in the IVCs is observed down to 1.3 K. This is anticipated from the model discussed earlier. As $L$ is increased to about 1.4 $\mu$H, large $V(\Phi)$ oscillations are obtained over a wide range of bias current, see \mbox{Figs. \ref{fig:bvosc1}}(a,c). IVCs at 1.3 K and different flux values, shown in \mbox{Fig. \ref{fig:bvosc1}(d)}, display complete reversibility and smooth transitions in contrast to irreversible and sharp switches observed without inductive shunt in \mbox{Fig. \ref{fig:ivc}(a)}. The dynamic retrapping current $I_{\rm r}^{\rm dyn}$ matches the critical current $I_{\rm c}^{\rm 0}$. The latter does not change, as expected.

The dashed line in \mbox{Fig. \ref{fig:bvosc1}(b)} shows the best fit of the zero-field IVC at 1.3 K to the DTM, showing two transitions at $I_{\rm c}^{\rm 0}$ and $I_{\rm r}^{\rm dyn}$. The fit is good, given that the model does not include the effect of fluctuations arising from thermal and other extrinsic effects, which lead to rounding in IVCs when $I_{\rm r}^{\rm dyn}$ and $I_{\rm c}^{\rm 0}$ are close.\cite{squidbook,nikhilsust,sourav} We take $\beta$(1.3 K) = 9.3 from \mbox{Fig. \ref{fig:ivc}(a)} IVC fitting and the given $r$ = 4.2. The single fit parameter $\gamma/\alpha$ is found to be about 1.01, which gives $\tau_{\rm th} \approx$ 0.8 $\mu$s. Using $k$ = 4.3 nW/K (see Appendix B), the effective heat capacity $C_{\rm WL}$ is then estimated to be 3.4 $\times$ 10$^{\rm -15}$ J/K. Based on the tabulated \cite{Nb specific heat} specific heat of 25.7$\times$10$^{-3}$ J/cc.K of Nb just below $T_{\rm c}$, we obtain a volume of $13\times10^{-2}$ $\mu$m$^3$, \mbox{i.e.} a film surface of 6.5 $\mu$m$^2$. Therefore, the heat generation in the dissipative state of each WL happens over an effective area of 3.25 $\mu$m$^2$, which is well above the mere WL area of 64$\times$10$^{-4}$ $\mu$m$^2$. Earlier experiments \cite{dolan-PS,SBT-phase-slip} on WLs show that the Joule heat is indeed generated over a length scale determined by the inelastic quasiparticle diffusion length. The obtained thermal time $\tau_{\rm th}$ agrees well with the typical quasiparticle recombination time in Nb.\cite{kaplan,x-ray} Thus the real bottleneck in healing back the superconductivity in the WL is not the heat evacuation from the phonons. It is rather the slow recombination of quasiparticles, which ensure the energy transfer to phonons.\cite{cooler,qbit}

\begin{figure}[t!]
	\includegraphics[width=0.94\columnwidth]{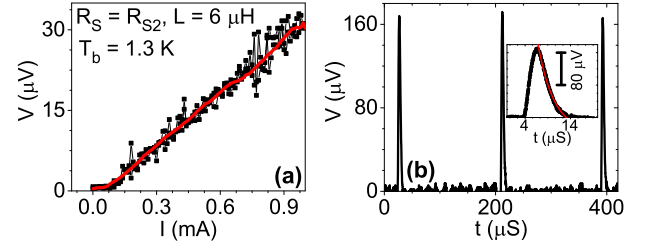}
	\caption{(a) IVCs in the relaxation regime for the device with a shunt $L$ = 6 $\mu$H at 1.3 K. (b) Voltage signal with time at a bias current of 140 $\mu$A. Inset: One zoomed peak showing relaxation oscillation in voltage. Red line is the fit to an exponential decay function.}
	\label{fig:relax}
\end{figure}

At a bath temperature $T_{\rm b}$ = 1.3 K and at the optimal bias, the flux-to-voltage transduction function $V_\Phi$=$\mid\partial V/\partial\Phi(\Phi)\mid_{\rm max}$ is found to be 680 $\mu$V/$\Phi_{\rm 0}$. Thus we obtain a flux noise density $\sqrt{S_\Phi} \simeq$ 6 $\mu \Phi_{\rm 0}/\sqrt{\rm Hz}$. Here we use the estimated voltage white noise (above 100 Hz) in our amplifier as 4 nV$/\sqrt{\rm Hz}$. The corresponding white-noise limited spin sensitivity, defined by $\sqrt{S_{\rm n}}=\sqrt{S_\Phi}/\Phi_{\mu}$, is estimated to be 10$^3$ $\mu_{\rm B}/\sqrt{Hz}$. The coupling factor writes \cite{granatasquid} $\Phi_{\mu}=2\sqrt{2}\mu_0\mu_{\rm B}/\pi \mathcal{L}$ with $\mathcal{L}$ the side length of the SQUID loop. At a higher bath temperature $T_{\rm b}$ = 1.6 K, the voltage modulation amplitudes are smaller but the transduction function $V_\Phi$ increases significantly to 2.45 mV$/\Phi_{\rm 0}$, see \mbox{Fig. \ref{fig:bvosc1}(c)}. In this case, a very low flux noise density $\sqrt{S_\Phi}\sim$ 1.6 $\mu\Phi_{\rm 0}/\sqrt{Hz}$, corresponding to a spin sensitivity $\sqrt{S_{\rm n}}\sim$ 300 $\mu_{\rm B}/\sqrt{Hz}$, is achieved. This figure could be further improved by using a low temperature amplifier with lower voltage noise.

Based on the model, the relaxation oscillation regime in IVCs is expected to start above $\gamma/\alpha=1.15$, \mbox{i.e.} $L\approx$ 1.84 $\mu$H at 1.3 K. At a somewhat higher value of $L$ = 6 $\mu$H, \mbox{i.e.} $\gamma/\alpha=3.75$, clear relaxation oscillations in voltage are observed for a fixed current bias, as seen in Fig. \ref{fig:relax}(b). Depending on time averaging and sampling rate, the IVC in this regime carry excess noise, as seen in Fig. \ref{fig:relax}(a). The fit of the decay part of the voltage peak to an exponential gives a time constant of 3.53 $\mu$s, which matches well with the calculated $\tau_{\rm L}=L/R_{\rm S2}=$ 3.6 $\mu$s. The relaxation oscillations in Josephson junctions have been extensively studied with an understanding based on either static thermal models or the RCSJ model.\cite{karl,whan1,vernon}

\begin{figure}[t!]
	\includegraphics[width=0.6\columnwidth]{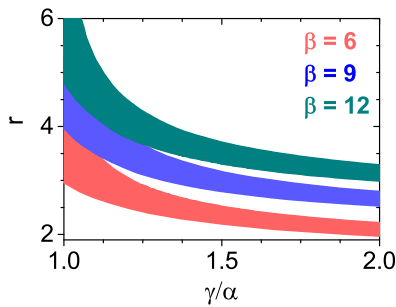}
	\caption{Domains in $r-\gamma/\alpha$ space for three $\beta$ values showing the most suitable shunt parameters. Upper and lower border of a domain correspond to $i_{\rm r}^{\rm dyn}=$ 1 and 0.95, respectively.}
	\label{fig:phase}
\end{figure}

The shunt inductance is thus found to be an important parameter that directly controls the current, phase and temperature dynamics in the WLs of a shunted $\mu$-SQUID. It is the relative magnitude of $\tau_{\rm th}$ and $\tau_{\rm L}$ that determines the physics of the WL. In order to get a reversible non-hysteretic regime where the dynamic retrapping current $I_{\rm r}^{\rm dyn}$ is close to the critical current $I_{\rm c}^{\rm 0}$, the inductance value needs to be adjusted so that $\tau_{\rm L}$ is of the same order as $\tau_{\rm th}$. Figure \ref{fig:phase} shows the region in $r-\gamma/\alpha$ space in which a $\mu$-SQUID would be practically reversible and useful for flux-to-voltage transducer at low temperature (higher $\beta$). The regions above and below this region give relaxation oscillations and hysteretic IVCs, respectively.

\section{Conclusion}
In conclusion, we have discovered that shunting a superconducting WL with a fine-tuned inductance can eliminate thermal hysteresis and provide a large voltage modulation by the magnetic flux in a $\mu$-SQUID well below the critical temperature. This result is opposed to the usual belief that an inductive shunt gives rise to relaxation oscillations. Such inductive shunts could be realized with disordered superconductors featuring a high-kinetic inductance.\cite{pop-kin-ind} While the consistent fabrication of fully nonhysteretic $\mu$-SQUIDs at all temperatures is still a challenge, this study demonstrates a practical procedure for getting a reliable voltage read-out of the flux using usual hysteretic $\mu$-SQUIDs, which opens up an easy way for using such devices for nanoscale magnetism, in particular at very low temperatures.

\section*{ACKNOWLEDGMENTS}
We are indebted to T. Crozes for help in the device fabrication at the Nanofab platform at N\'eel Institute. SB acknowledges a financial grant from CSIR, Government of India and IIT Kanpur. AKG acknowledges a research grant from the SERB-DST of the Government of India. AKG thanks Universit\'e Grenoble Alpes for an invited professorship. CBW and HC acknowledge financial support from the LabEx LANEF project (ANR-10-LABX-51-01), and we acknowledge a research grant (5804-2) from CEFIPRA.

\section*{APPENDIX A: ADDITIONAL RESULTS ON THE MODEL}
\begin{figure}[t!]
	\centering
	\includegraphics[width=\columnwidth]{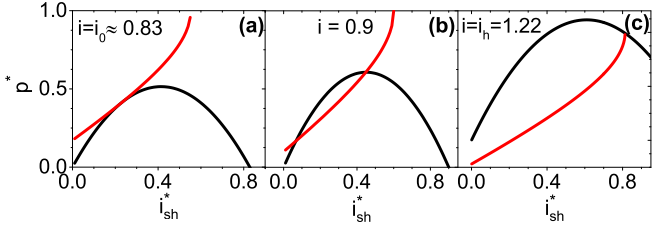}
	\caption{Plots of Eq. (\ref{eq:eq6}) (black) and Eq. (\ref{eq:eq7}) (red) for different $i$ values with $\beta=6$ and $r=2$ at (a) $i=i_0\approx0.83$, \mbox{i.e.} the bifurcation point, (b) at $i=0.9$ and (c) at upper limit of $i$, \mbox{i.e.} $i=i_h\approx1.22$ [see Fig.\ref{fig:model2}(a)], for the dynamic regime.}
	\label{fig:fixedpt}
\end{figure}

Using \mbox{Eqs. (\ref{eq:eq6}) and (\ref{eq:eq7})} for the fixed point coordinates, $p^*$ and $i_{\rm sh}^*$, one can eliminate $p^*$ to obtain a quartic equation in $i_{\rm sh}^*$:
\begin{equation}a\,{i_{\rm sh}^*}^4+b\,{i_{\rm sh}^*}^3+c\,{i_{\rm sh}^*}^2+d\,{i_{\rm sh}^*}+e=0
\label{eq:eq11}.
\end{equation}
Here, $a=\beta^2$, $b=-2\beta^2i$, $c=1-r^2+2\beta r+\beta^2i^2$, $d=2ir^2-2\beta ir$ and $e=(1-i^2)r^2$. However, the formula with such coefficients is unwieldy to get an analytical expression for $i_{\rm sh}^*$ in terms of the parameters $\beta$, $i$ and $r$. A more insightful approach is to plot $p^*$ and $i_{\rm sh}^*$ as per \mbox{Eqs. (\ref{eq:eq6}) and (\ref{eq:eq7})} as shown in \mbox{Fig. \ref{fig:fixedpt}}(a-c) for $\beta=6$ and $r=2$ at different $i$ values. The intersection points of the two curves give possible fixed point coordinates ($p^*$, $i_{\rm sh}^*$). These have been numerically computed as a function of $i$ and plotted in \mbox{Fig. \ref{fig:model1}(a)}.

At the bifurcation point $i=i_0$, the two functions are tangent at the only coinciding point, see \mbox{Fig. \ref{fig:fixedpt}(a)}. The value of $i_0$ for a given $\beta$ and $r$ is computed from the condition of two equal roots of the Eq. (\ref{eq:eq11}). Below $i_0$, there is no intersection and hence no dynamic steady state is possible. Above $i=i_0$, there are two intersections: one at relatively small values of $p^*$ and $i_{\rm sh}^*$ and the other at larger values, see \mbox{Fig. \ref{fig:fixedpt}(b)}.
The Jacobian matrix ($J$) associated with the dynamical system given by \mbox{Eqs. (\ref{eq:eq4}) and (\ref{eq:eq5})}, at the fixed point ($p^*,i_{\rm sh}^*$), works out as
\begin{widetext}
\begin{align}
J & =
\begin{pmatrix}
\frac{\partial \langle\dot{p}\rangle}{\partial \overline{p}} & 	\frac{\partial \langle\dot{p}\rangle}{\partial \overline{i}_{\rm sh}}\\ \\
\frac{\partial \langle\dot{i_{\rm sh}}\rangle}{\partial \overline{p}} & 	\frac{\partial \langle\dot{i_{\rm sh}}\rangle}{\partial \overline{i}_{\rm sh}}
\end{pmatrix}_{(p^*,i_{\rm sh}^*)}=
\begin{pmatrix}
-\frac{\gamma}{\alpha}+\frac{\beta r \gamma}{\alpha}\frac{(i-i_{\rm sh}^*)(1-p^*)}{i_{\rm sh}^*} & 	-\frac{\beta r \gamma}{\alpha}\frac{2(i-i_{\rm sh}^*)^2-(1-p^*)^2}{i_{\rm sh}^*}\\
r^2\frac{1-p^*}{i_{\rm sh}^*} & -1-r^2\frac{i-i_{\rm sh}^*}{i_{\rm sh}^*}
\end{pmatrix}\label{eq:jacobian2}.
\end{align}
\end{widetext}	

\begin{figure}[t]
	\includegraphics[width=0.6\columnwidth]{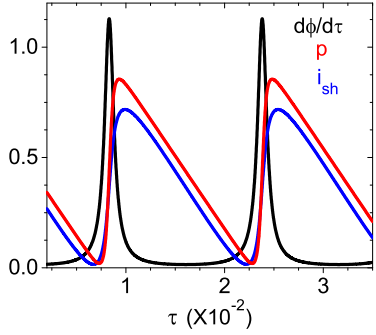}
	\caption{Small-scale zoom-in of the $p$ [actually $300(p-0.4413)$], $i_{\rm sh}$ [actually $300(i_{\rm sh}-0.2568)$] and $\dot{\phi}$ time traces at $i=0.83$ for $\gamma/\alpha=0.5$.}
	\label{fig:model_suppl}
\end{figure}
The stability of the fixed points is obtained from the trace ($\Tr$) and determinant ($\Delta$) of $J$ given in Eqs. (\ref{eq:eq8}) and (\ref{eq:eq9}). The evolution of the nature of the fixed point can be better illustrated using a vector flow diagram.\cite{sourav-thesis}


The small steady oscillations in temperature $p$ and $i_{\rm sh}$ around their average values ($p^*,i_{\rm sh}^*$) are shown along with $\dot{\phi}$ oscillation in \mbox{Fig. \ref{fig:model_suppl}}. The dynamic steady state exists till a certain bias current, called static retrapping current $i_{\rm h}$, at which $p^*$ reaches 1.\cite{sourav} Putting $p^*=1$ in \mbox{Eqs. (\ref{eq:eq6}) and (\ref{eq:eq7})}, the formula for $i_{\rm h}$ expectedly comes out to be $(1+r)/\sqrt{\beta}$. Note that $i_{\rm h}$ is independent of $\gamma/\alpha$ unlike $i_{\rm r}^{\rm dyn}$.

\section*{APPENDIX B: ADDITIONAL RESULTS ON PURE RESISTIVE SHUNTING}

\begin{figure}[t!]
	\includegraphics[width=0.75\columnwidth]{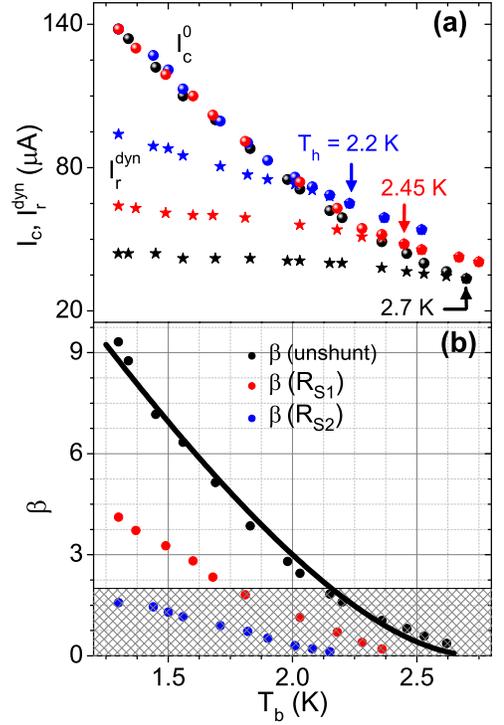}
	\caption{(a) Dependence of $I_{\rm c}^0$ and $I_{\rm r}^{\rm dyn}$ on the bath temperature $T_{\rm b}$ for a resistively-shunted device (no inductance). (b) $\beta$ variation with $T_{\rm b} (<T_{\rm h})$ calculated from the ratio $I_{\rm r}^{\rm dyn}/I_{\rm c}^{\rm 0}$ and using $I_{\rm r}^{\rm dyn}$ expression.\cite{sourav} Solid line is the fit to the DTM for unshunted device. The shaded area in panel (b) depicts the parameter range $\beta\leq2$ where $V(\Phi)$ oscillations are significant.}
\label{fig:schematic}
\end{figure}

The bath temperature $T_{\rm b}$ dependence of $I_{\rm c}^{\rm 0}$ and $I_{\rm r}^{\rm dyn}$ is shown in \mbox{Fig. \ref{fig:schematic}(a)} when the shunt inductance is negligible. The crossover between the reversible ($I_{\rm r}^{\rm dyn}\approx I_{\rm c}^{\rm 0}$) and hysteretic ($I_{\rm r}^{\rm dyn} < I_{\rm c}^{\rm 0}$) regimes occurs at a temperature $T_{\rm h}$ that decreases slightly by incorporating a shunt. This, together with the slightly larger $I_{\rm c}^{\rm 0}$ above $T_{\rm h}$ in shunted devices is attributed to the distribution of current fluctuations between the shunt and the WLs, leading to a decrease in the WLs heating and thus an increase in $I_{\rm c}^{\rm 0}$ value.\cite{bezryadin} For zero-inductance limit, in the hysteretic regime, we obtain the $\beta$ parameter value as a function of $T_{\rm b}$ from the measurement of the $I_{\rm r}^{\rm dyn}$ value and using $I_{\rm r}^{\rm dyn}$ expression,\cite{sourav,anjanjap} see \mbox{Fig. \ref{fig:schematic}(b)}. Similar to our earlier analysis on unshunted $\mu$-SQUIDs, the fit of $\beta$ variation to the DTM gives $k=4.3$ nW/K.

\begin{figure}[t!]
	\includegraphics[width=0.9\columnwidth]{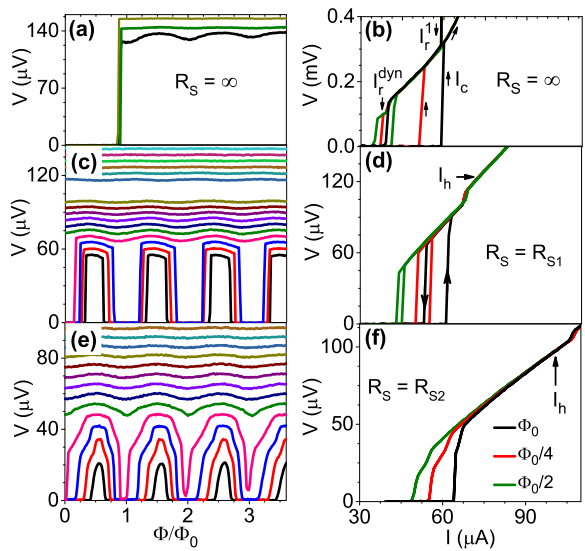}
	\caption{$V(\Phi)$ modulations and IVCs at different flux values ($0, \Phi_{\rm 0}/2$ and $\Phi_{\rm 0}/4$) at 2.2 K for no shunt, a $R_{\rm S1}$ shunt and a $R_{\rm S2}$ shunt. Bias current ranges for V-$\Phi$ modulations are 40-42 $\mu$A, 48-80 $\mu$A and 52-100 $\mu$A for $R_{\rm S}=\infty$, $R_{\rm S1}$ and $R_{\rm S2}$ respectively.}
	\label{fig:BV}
\end{figure}

The voltage modulation by the flux for the three shunt cases is displayed in \mbox{Figs. \ref{fig:BV}(a,c,e)} for a bath temperature $T_{\rm b}$ = 2.2 K. The periodicity in magnetic field is consistent with a flux $\Phi_{\rm 0}$ over an effective SQUID loop area of 1.8 $\mu$m$^2$. In the unshunted device with higher $\beta$, voltage oscillations are expectedly seen only over a short bias current range just above $I_{\rm r}^{\rm dyn}$. The shunted devices display voltage oscillations over a larger bias current range. The IVCs show a consistent behavior with the $V(\Phi)$ data, see \mbox{Figs. \ref{fig:BV}(b,d,f)}, the $R_{\rm S2}$-shunted device being nonhysteretic. At further lower temperature, the parameter $\beta_{\rm 0}$ being higher, no voltage modulation could be observed in the unshunted case. However, oscillations are seen for the $R_{\rm S1}$ shunt till 1.8 K. The dynamic regime becomes wider for $R_{\rm S2}$-shunted device, although the IVCs remain hysteretic.\cite{sourav-thesis}

The flux-to-voltage transduction function $V_\Phi$=$\mid\partial V/\partial \Phi(\Phi)\mid_{\rm max}$ is found to be 40 $\mu$V/$\Phi_{\rm 0}$ for the unshunted device at 2.2 K just above $I_{\rm r}^{\rm dyn}$, which leads to a flux noise density $\sqrt{S_{\Phi}}$ = 100 $\mu \Phi_{\rm 0}/\sqrt{\rm Hz}$ for the voltage white noise of 4 nV$/\sqrt{\rm Hz}$. The IVCs being nonhysteretic for $R_{\rm S2}$ shunt at 2.2 K, $V_\Phi$ increases to 132 $\mu$V/$\Phi_{\rm 0}$. Thus, we get a reduced $\sqrt{S_{\Phi}}$ = 30 $\mu \Phi_{\rm 0}/\sqrt{\rm Hz}$ with the $R_{\rm S2}$ shunt.

\end{document}